\documentclass[notitlepage,prl,twocolumn,showpacs,preprintnumbers,amsmath,nofootinbib,amssymb,superscriptaddress, hyperref]{revtex4-1}
\pdfoutput=1
\usepackage{amssymb,amsmath,latexsym,mathrsfs}
\usepackage[usenames,dvipsnames]{color}
\usepackage{graphicx} 
\usepackage{bm}
\usepackage{natbib}
\usepackage{todonotes}
\usepackage{color}
\usepackage[normalem]{ulem}
\usepackage[breaklinks,colorlinks,urlcolor=blue,citecolor=blue,linkcolor=blue]{hyperref}
\usepackage{epsfig,graphicx,verbatim,xspace,multirow,mathtools}
\usepackage{array}

\def\gsim{\mathrel{\raise.3ex\hbox{$>$\kern-.75em\lower1ex\hbox{$\sim$}}}}

\newcommand{\Fig}[1]{Fig.~\ref{#1}}
\newcommand{\Eq}[1]{Eq.~\ref{#1}}
\newcommand{\ie}{\emph{i.e.~}}
\newcommand{\eg}{\emph{e.g.~}}

\begin{document}

\title{Looking for Axion Dark Matter in Dwarf Spheroidals}
\author{Andrea Caputo}
\affiliation{Instituto de F\'{\i}sica Corpuscular (IFIC), CSIC-Universitat de Val\`encia, \\
Apartado de Correos 22085,  E-46071, Spain}
\author{Carlos Pe\~{n}a Garay}
\affiliation{I2SysBio, CSIC-UVEG, P.O.  22085, Valencia, 46071, Spain}
\affiliation{Laboratorio Subterr\'aneo de Canfranc, Estaci\'on de Canfranc, 22880, Spain}
\author{Samuel J.~Witte}
	\affiliation{Instituto de F\'{\i}sica Corpuscular (IFIC), CSIC-Universitat de Val\`encia, \\
Apartado de Correos 22085,  E-46071, Spain}

\begin{abstract}

We study the extent to which the decay of cold dark matter axions can be probed with forthcoming radio telescopes such as the Square Kilometer Array (SKA). In particular we focus on signals arising from dwarf spheroidal galaxies, where astrophysical uncertainties are reduced and the expected magnetic field strengths are such that signals arising from axion decay may dominate over axion-photon conversion in a magnetic field. We show that with $\sim100$ hours of observing time, SKA could improve current sensitivity by a factor of about five.

\end{abstract}
%%%%%%%%%%%%%%%%%%%%%%%%%%%%%%%%%%%%%%%%%%%%%%%%%%%%%
\maketitle
%%%%%%%%%%%%%%%%%%%%%%%%%%%%%%%%%%%%%%%%%%%%%%%%%%%%%
%\newpage

\section{Introduction}

Perhaps the most promising solution to the strong CP problem involves the introduction of a new global $U(1)$ symmetry, which when spontaneously broken at high energies produces a pseudo-Nambu-Goldstone boson known as the `axion'~\cite{Peccei:1977hh,Peccei:1977ur,Weinberg:1977ma,Wilczek:1977pj} (see \eg\cite{Peccei:2006as} for a recent review). Despite the small mass of the axion, non-thermal production mechanisms in the early Universe allow for the possibility that this particle accounts for the observed abundance of cold dark matter~\cite{Preskill:1982cy,Dine:1982ah,Abbott:1982af,Davis:1986xc,Lyth:1991bb}\footnote{It is worth noting that there has been a recent interest in the community in searching for axion-like particles (ALPs); in such models one sacrifices the ability to solve the strong CP problem for the freedom to independently choose the mass and coupling of the pseudoscalar~\cite{Irastorza:2018dyq}.}.

The potential of simultaneously solving two of the largest outstanding problems in particle physics has lead to a large and diverse experimental program aimed at probing the parameter space associated with axion dark matter (see \eg\cite{Irastorza:2018dyq} for a review of experimental search techniques). Many of these experiments have been focused on exploiting the conversion of axions to and from photons, a process which occurs in the presence of a strong magnetic field; this is the so-called Primakoff Effect. In this work, we instead focus on the often neglected process of the spontaneous decay of the axion into two photons, and show that near-future radio telescopes can significantly improve existing sensitivity to axions with masses $m_a \sim 10^{-6}-10^{-4}$ eV.

Here, we focus on the signal arising from on one particular class of astrophysical objects, dwarf spheroidal galaxies (dSphs) -- these are dark matter dominated objects with low radio background, and are thus capable of yielding significant unabated signatures of axion dark matter. Conventional astrophysical searches for axion conversion (\eg targeting the Galactic Center~\cite{Kelley:2017vaa} or highly magnetized pulsars~\cite{Huang:2018lxq,Hook:2018iia}), on the other hand, typically require targeting highly uncertain environments with many assumptions on \eg the magnetic field strength, the magnetic field distribution, radio backgrounds arising from synchrotron emission and free-free absorption, and the dark matter distribution. 

The possibility of detecting axion decay in dSphs with a radio telescope was first explored in~\cite{Blout:2000uc}, and to our knowledge, not discussed since. The subsequent $\sim20$ years have provided significant improvement in the sensitivity of radio telescopes and the discovery of many new dSphs. In this work we evaluate the sensitivity of current and future radio telescopes to axion decay in a variety of known dSphs. In particular, we show that with $\sim 100$ hours of observation, the Square Kilometer Array (SKA)~\cite{WinNT} can potentially improve the constraints from helioscopes by half an order of magnitude depending on the axion mass. For completeness, we also present a comparison of the signal arising from axion decay with that produced by the process of magnetic field conversion.   

\begin{figure}
	\includegraphics[width=0.45\textwidth]{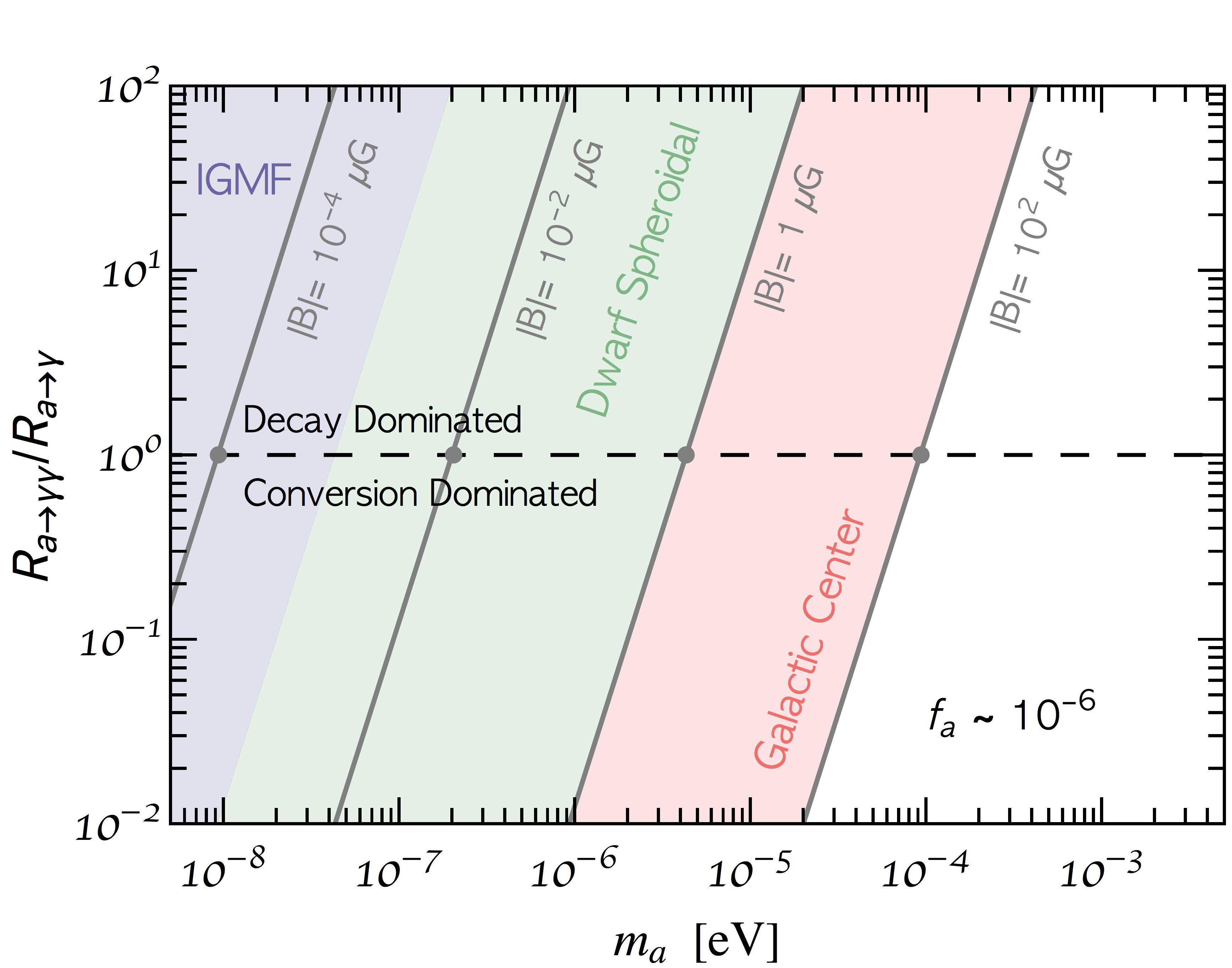}
	\caption{\label{comparison}Ratio of the rate of spontaneous decay $a \rightarrow \gamma\gamma$ to the conversion $a\rightarrow\gamma$ in a spatially uniform magnetic field of modulus $|B|$ as a function of the axion mass. While magnetic field conversion is far more efficient at small axion masses, the rate of spontaneous decay can be competitive for $m_a \sim 10^{-5}eV$, depending on the value of the magnetic field. Approximate magnetic field strengths for the Galactic Center (red), dSphs (green), and the intergalactic medium (grey) are also shown for comparison. Here, we have taken $f_a =10^{-6}$, which is likely extremely optimistic for many large scale enviornments.    }
\end{figure}

\section{The signal}
\subsection{Axion Decay}
The spontaneous decay of an axion with mass $m_a$ proceeds through the chiral anomaly to two photons, each with a frequency $\nu = m_a/4\pi$, and with a lifetime given by 
\begin{equation}\label{decay}
\tau_a = \frac{64\pi}{m_a^3 g^2} \, ,
\end{equation}
where $g$ is the axion-to-two-photon coupling constant. It is perhaps straightforward to understand why axion decay has been largely neglected in the literature -- if one evaluates \Eq{decay} for $m_a \sim 10^{-4}$ eV and $g \sim 7\times 10^{-11}$ (\ie the upper limit imposed by helioscopes), it is found that the axion has a lifetime of $\sim 10^{30}$ years.

The observed power per unit area per unit frequency from the spontaneous decay of an axion by a radio telescope is then given by \begin{equation}\label{flux}
S_{\rm sd} = \frac{m_a}{4\pi\Delta\nu}\int \, d\Omega \, d\ell \, \frac{n(\ell,\Omega)}{\tau_a} \, ,
\end{equation}
where $\Delta\nu$ is the width of the axion line, given by $\Delta\nu = \nu_{\rm center}\sigma_{disp}$ where $\sigma_{disp}$ is the dispersion of the line resulting from the velocity dispersion of dark matter within the dwarf\footnote{Note that we assume here that the velocity dispersion of dark matter follows that of the stellar component.}, and $n(\ell, \Omega)$ is the axion number density at distance along the line of sight $\ell$ in the solid angle $\Omega$. For convenience, we define the astrophysical quantity 
\begin{equation}\label{eq:dfac}
D(\alpha_{\rm int})=\int d\Omega \, d\ell \, \rho(\ell,\Omega) \, ,
\end{equation}
which allows us to write \Eq{flux} as
\begin{equation}\label{eq:power2}
S_{\rm sd} = \frac{m_a^2 g^2}{64 \pi } \, \frac{D(\alpha_{\rm int})}{\sigma_{\rm disp}} \, ,
\end{equation}
where $\alpha_{\rm int}$ is the angle of integration defined between the center of a given dwarf galaxy and the largest observable radius. From \Eq{eq:power2} it should be clear that given a radio telescope, the dwarf galaxies providing the best sensitivity to axion decay are those with the largest $D(\alpha_{\rm int})/\sigma_{\rm disp}$. The angle $\alpha_{\rm int}$ depends on both the size of the dishes used for observation and the frequency (and thus the axion mass). The observed field-of-view of a particular telescope is given by    
\begin{equation}
\Omega_{FoV} \sim \frac{\pi}{4}\left(\frac{66\lambda}{D_{dish}}\right)^2 \, , 
\end{equation}
where $\lambda$ is the wavelength of the observed photons and $D_{dish}$ is the diameter of the dish. Consequently, the maximum angle of the dwarf that can be probed is given by
\begin{equation}\label{eq:alphaint}
\alpha_{\rm int} \sim 8.8 \times \left(\frac{{\rm GHz}}{\nu}\right)\left(\frac{1 {\rm m}}{D_{dish}}\right) \, \hspace{.1cm} {\rm degrees} .
\end{equation}

Up until this point we have focused solely on the spontaneous decay of axions into photons. However, these decays take place in background of CMB photons with the same energy; this inherently enhances the photon production rate via stimulated emission. The power emitted by stimulated emission is given by 
\begin{equation}
P_{se} = m_a g(\nu) B_{21} \rho_{cmb}(\nu)\Delta N \, ,
\end{equation}
where $B_{21}$ is the Einstein coefficient, $\rho_{cmb}(\nu)$ is the radiation density of CMB photons at a given frequency, $\Delta N$ is the difference between the number of axions and the number of photons with energy $m_a/2$ (which can be approximated as just $N_a$), and $g(\nu)$ is the spectral line shape function (which we take to be a delta function). Taking the Einstein coefficient to be
\begin{equation}
B_{21} = \left(\frac{\pi^2}{\omega^3}\right)\frac{1}{\tau} = \frac{g^2 \pi}{8} \, ,
\end{equation}
one can express the observed power per unit area per unit frequency contributed from stimulated emission as
\begin{equation}\label{eq:se}
S_{\rm se} = 2\,\frac{m_a^2 g^2}{64\pi} \, \frac{D(\alpha_{\rm int})}{\sigma_{disp}} \left( \frac{1}{e^{m_a/(2T_{cmb})} - 1} \right) \, ,
\end{equation}
which is identical to the contribution of the spontaneous emission multiplied by two times the photon occupation number. We emphasize that the aforementioned effect of stimulated emission, described above using the language of atomic physics, is entirely equivalent to the QFT computation performed assuming a background of photons rather than a vacuum. An identical result can be obtained using the Boltzmann equation and applying the limit in which the photon number density is much less than the axion number density. The total observed power $S_{\rm total}$ is then given by the sum of \Eq{eq:power2} and \Eq{eq:se}.

\subsection{Magnetic Conversion}

Signals arising from the conversion of axions to photons while traversing through a perpendicular magnetic field has been extensively studied in the context of both terrestrial and astrophysical environments (see \eg\cite{Shokair:2014rna,Budker:2013hfa,Brubaker:2016ktl,Kenany:2016tta,Majorovits:2016yvk,Brubaker:2017rna,Kelley:2017vaa,Huang:2018lxq,Hook:2018iia}). In this section we will make a number of simplifying assumptions to estimate the potential signal arising from axion conversion in a magnetic field under rather optimistic assumptions. As shown below, the rate of magnetic field conversion in large astrophysical environments is almost certainly subdominant to the contribution from axion decay, largely due to the suppression arising from the fact that realistic magnetic fields are not homogenous.

In the case of a static magnetic field, the flux arising from axion conversion is proportional to $|\vec{B}(|k| = m_a)|^2$, where $\vec{B}(\vec{k})$ is the Fourier transform of the magnetic field. It is often convenient to characterize this contribution in terms of a characteristic magnetic field strength $B_0$ and a suppression factor $f(m_a)$, as $|\vec{B}(|k| = m_a)|^2 = B_0^2 f(m_a)$ (see \eg Eq.~20 of~\cite{Sigl:2017sew}). Using this simplification, one can calculate the observed power per unit area per unit frequency from axion-photon conversion using~\cite{Sikivie:1983ip,Sigl:2017sew,Irastorza:2018dyq}:
\begin{equation}\label{eq:convert}
S_{conv} =  \frac{B_0^2 g^2 }{m_a^2 \sigma_{\rm disp}} \, f(m_a) \, \int_{V_{B_0}} \, d\Omega\, d\ell \, \rho(\ell,\Omega) \, ,
\end{equation}
where the integration runs over the volume containing the magnetic field, $V_{B_0}$. Naively, one may expect the astrophysical integral in \Eq{eq:convert} to be comparable to that of \Eq{eq:dfac}. Should this be the case, one may approximate the ratio between the rate of axion decay and the rate of axion conversion as
\begin{equation} \label{eq:ratecomp}
\frac{R_{a\rightarrow\gamma\gamma}}{R_{a\rightarrow\gamma}}\sim \frac{m_a^4}{64\pi B_0^2\, f(m_a)}\left(1+\frac{2}{e^{m_a/(2T_{cmb})} - 1} \right) \, .
\end{equation}
In \Fig{comparison} we plot \Eq{eq:ratecomp}, taking $f(m_a) \sim 10^{-6}$ (a value that will be motivated momentarily as extremely optimistic), as a function of the axion mass for various magnetic field strengths ranging from $100 \mu$G to $10^{-4}\mu$G. For comparison, we have also highlighted in \Fig{comparison} the approximate magnetic field strength expected in the galactic center~\cite{LaRosa2006,Beck:2008ty,Ferriere:2009dh,Han:2009ts,Beck:2013bxa}, dSphs~\cite{Regis:2014koa}, and the intergalactic medium~\cite{Arlen:2012iy,Finke:2015ona}. This first order comparison between the rate of decay and magnetic field conversion clearly illustrates an important point: the rate of axion decay can supersede that of magnetic field conversion for axion masses probed by radio telescopes.

At this point we emphasize that the suppression factor for large-scale astrophysical environments is likely $f(m_a) \ll 1$. For example, it was shown in \cite{Sigl:2017sew} that the rate of axion conversion in the Galactic Center is generically is reduced by a factor of $f(m_a) \sim 10^{-13}$, assuming the magnetic field distribution follows a power law and the coherence length in the Galactic Center is of order $\sim 1$ pc. Thus the adopted value of $f(m_a)$ in \Fig{comparison} of $10^{-6}$ is likely incredibly optimistic, implying decay will supersede that of conversion in most large scale environments. While the coherence length of the magnetic field in dSphs is likely considerably smaller than that of the Galactic Center, the length scales are still astrophysical and thus should reduce the overall rate of conversion by orders of magnitude. Thus it should be clear that axion decay, and not magnetic field conversion, is the relevant mechanism for the detection of axions in large-scale astrophysical environments.

\subsection{Sensitivity}

It is often conventional in radio astronomy to define the brightness temperature induced by a flux $S_{\rm total}$ as
\begin{equation}\label{eq:tbright}
T=\frac{A_{\rm eff}S_{\rm total}}{2} \, ,
\end{equation}
where $A_{\rm eff}$ is the effective collecting area of the telescope. Provided astrophysical backgrounds are low, the minimum observable temperature in a bandwidth $\Delta B$ given an observation time $t_{\rm obs}$ is given by 
\begin{equation}\label{eq:tmin}
T_{min}\sim\frac{T_{sys}}{\sqrt{\Delta B \times t_{\rm obs}\, N_{\rm tele} }}
\end{equation}
where $N_{\rm tele}$ is the number of telescopes and $T_{sys}$ consists of a sum over sky and instrumental noise, as well as residuals after background subtraction. For frequencies in the central part of the SKA band, it is typically valid to assume the sky temperature $T_{sky}$ is subdominant to the contribution from instrumental noise. Actually, because we are interested in searching for narrow spectral lines, the only backgrounds expected to be truly detrimental are molecular lines\footnote{{That is to say that spectrally smooth backgrounds are expected to be easily removed, implying the dominant background temperature for a majority of the frequency range considered is set by the instrumental noise.}}. However it seems reasonable to neglect the possibility that photons arising in axion decay directly coincide with the location of a molecular line. For completeness, our estimation of the system temperature includes both the quoted 11K instrumental noise and a frequency dependent contribution from the sky temperature~\cite{WinNT}. Throughout this analysis we will assume an observation time of 100 hours and a bandwidth such that at least two bins with the Nyquist width of the autocorrelation spectrometer are inside the axion signal~\cite{Blout:2000uc}, a feat that should be easily achievable for SKA (note that SKA is expected to have $\mathcal{O}({\rm kHz})$ sensitivity~\cite{SKABase}, far below any requirements imposed here).

Combining \Eq{eq:tbright} with \Eq{eq:tmin}, one can estimate the smallest detectable flux of a given radio telescope as
\begin{equation}
S_{\rm min} = 2\frac{T_{\rm sys}}{A_{\rm eff} \sqrt{ \Delta B \times t_{\rm obs} \, N_{\rm tele} }} \, .
\end{equation}

\section{Dwarf Galaxies}
In this analysis we focus on signals arising in dSphs, as these targets offer large dark matter densities and low velocity dispersion. DSphs also make an intriguing target for radio telescopes as the typical angular size of such objects is roughly the same order of magnitude as the field of view. Note that this is not inherently true of the Galactic Center, implying that radio telescopes can not take full advantage of the larger column density in this environment. It is also worth mentioning that targets like the Galactic Center have $(\emph{i})$ a larger velocity dispersion and $(\emph{ii})$ a larger magnetic field, where the former implies a suppression of the flux and the later an enhancement (large magnetic fields imply synchrotron emission, which can enhance the stimulated emission factor). These counteracting effects make it non-trivial to determine which environment is more promising. Here, we choose to focus on dSphs, and leave a more careful analysis of other environments to future work. 

\begin{figure}
\begin{center}
	\includegraphics[width=0.48\textwidth]{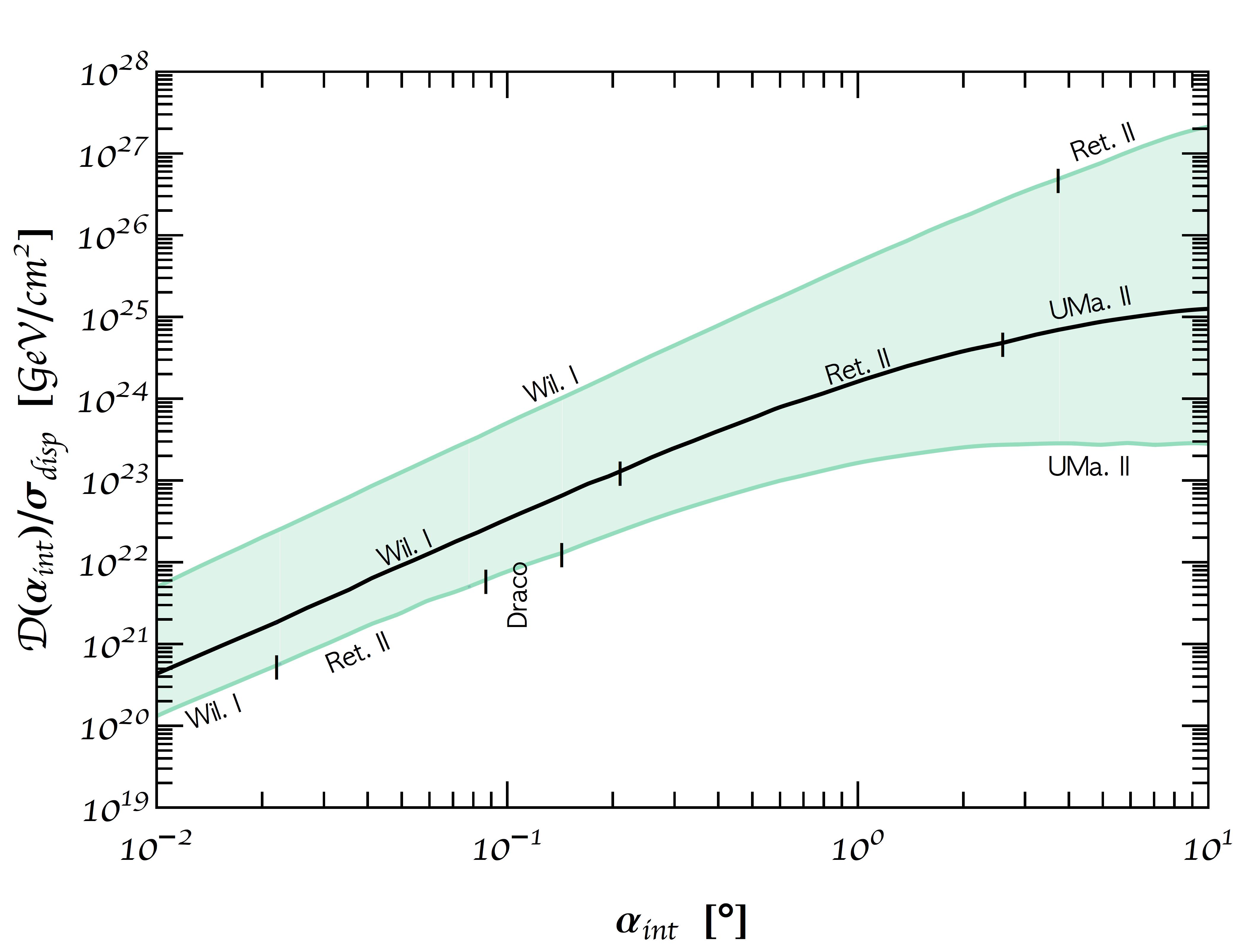}
	\caption{\label{Dwarf} Factor $D(\alpha_{\rm int})/\sigma_{\rm disp}$ as a function of the field of view of the telescope. For each value of $\alpha_{\rm int}$, we plot the median (black line) and $\pm 95\%$ CI (green band) of $D(\alpha_{\rm int})/\sigma_{\rm disp}$ for the dwarf galaxies providing the largest sensitivity. Depending on the field of view of a given radio telescope, the optimal dwarf galaxy is either Willman I, Reticulum II, Ursa Major II, or Draco.}
\end{center}
\label{Dwarfs}
\end{figure}

As previously stated, we restrict our analysis to only the most promising dSphs, \ie those producing the largest $D(\alpha_{\rm int})/\sigma_{disp}$. The value of $D(\alpha_{\rm int})$ for a wide variety of dSphs have been inferred from kinematic distributions (see \eg~\cite{Geringer-Sameth:2014yza,Bonnivard:2015xpq,Bonnivard:2015tta,Hayashi:2016kcy,Sanders:2016eie,Evans:2016xwx,Hayashi:2018uop,Petac:2018gue}). In this work we use the publicly available data provided in~\cite{Bonnivard:2015xpq,Bonnivard:2015tta} to determine the value of $D(\alpha_{\rm int})/\sigma_{disp}$ at each $\alpha_{\rm int}$. In \Fig{Dwarf} we show the maximum value of $D(\alpha_{\rm int})/\sigma_{disp}$ for all dSphs considered in~\cite{Bonnivard:2015xpq,Bonnivard:2015tta} as a function $\alpha_{\rm int}$, assuming the median best-fit $D(\alpha_{\rm int})$ (black line) and $\pm95\%$ CI $D(\alpha_{\rm int})$ (upper and lower green lines). The region bounded by the $\pm95\%$ CIs has been shaded to illustrate the range of possible values. For each line and each value of $\alpha_{\rm int}$, the dSph giving rise to the largest value of $D(\alpha_{\rm int})$ is identified. Note that the surprisingly continuous nature of the lines in \Fig{Dwarf} arises from the fact that a number of dwarfs considered have very similar values of $D(\alpha_{\rm int})/\sigma_{disp}$ (this is why we have chosen to mark the transitions with a vertical line segment). While the most promising dSph candidates are Reticulum II, Willman I\footnote{It is important to note that Willman I has irregular stellar kinematics indicative of a recent tidal disruption event, and thus the inferred distribution of dark matter may be untrustworthy.}, Ursa Major II, and Draco, we emphasize that many of the dSphs considered here produce values of $D(\alpha_{\rm int})/\sigma_{disp}$ comparable to those highlighted above; thus, the derived sensitivity is not specific to a particular choice of dSph.

\section{Results}
In this section we present the projected sensitivity for the SKA-Mid, operated in the phase-2 upgrade which assumes $\sim 5600$ telescopes, each with an effective area equal to the physical area of the single dish and a system temperature of $T_{\rm sys} = 11$K + $T_{sky}(\nu)$, where $T_{sky}(\nu)$ is taken from~\cite{WinNT}. We have also considered the sensitivity for various current and future radio telescopes, namely LOFAR, ASKAP, ParkesMB, WSRT, Arecibo, GBT, JVLA, FAST and SKA-Low, which will be presented elsewhere.

\begin{figure}
\begin{center}
\includegraphics[width=0.48\textwidth]{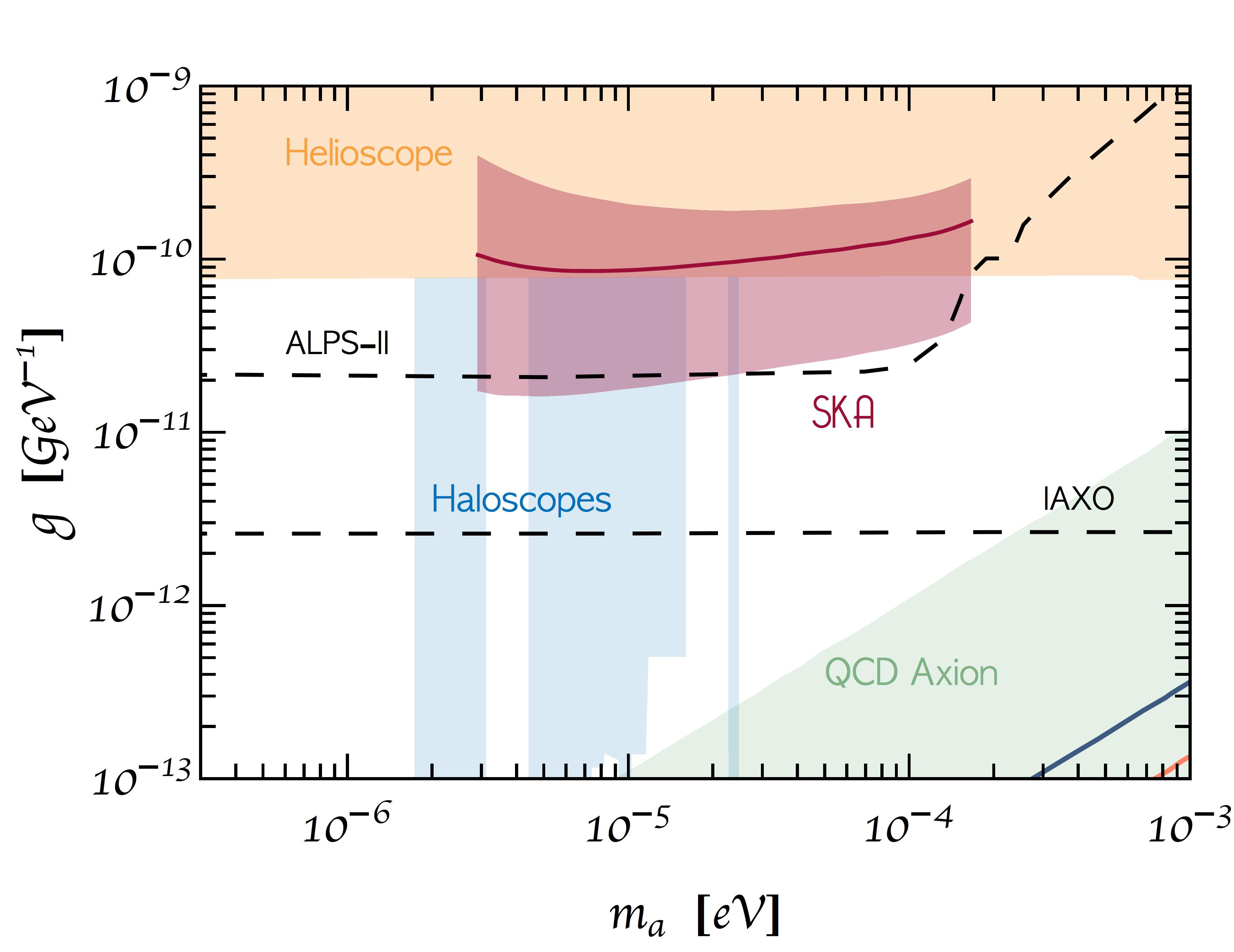}
\caption{\label{SKA} Sensitivity of SKA to the axion dark matter parameter space. The solid red line (bands) denote the median ($\pm 95\%$ CI) sensitivity to axion decay in the dSph providing the largest value of $D(\alpha_{\rm int})/\sigma_{\rm disp}$.  Sensitivity estimates are compared with the QCD axion parameter space \cite{DiLuzio:2016sbl}(light green), existing bounds from helioscopes~\cite{Anastassopoulos:2017ftl}, existing bounds from haloscopes~\cite{Hagmann:1990tj,Hagmann:1998cb,Asztalos:2001tf,Du:2018uak,Zhong:2018rsr}, and projected sensitivity for IAXO~\cite{Irastorza:2013dav} and ALPS-II~\cite{Bahre:2013ywa}. }
\end{center}
\end{figure}

\Fig{SKA} shows the estimated sensitivity for SKA-Mid (phase-2) to axion decay in dSphs. At each axion mass, the sensitivity curves shown in \Fig{SKA} are produced using the dSph providing the largest $D(\alpha_{\rm int})/\sigma_{\rm disp}$, where the width of the sensitivity band reflects the $\pm 95\%$ CL of $D(\alpha_{\rm int})$\footnote{One may wonder whether the sensitivity could be significantly improved by performing a stacked dSph analysis, as is done \eg in gamma-ray searches for dark matter. Unfortunately, the narrow field of view of radio telescopes implies they cannot simultaneously observe multiple dSphs, and thus a stacked analyses performed using a fixed total observation time will not result in an increased sensitivity, although it will reduce systematic uncertainties.}. This result is compared with current constraints from helioscopes~\cite{Anastassopoulos:2017ftl} and haloscopes~\cite{Hagmann:1990tj,Hagmann:1998cb,Asztalos:2001tf,Du:2018uak,Zhong:2018rsr,Melcon:2018dba}, projected sensitivity from future experiments IAXO~\cite{Irastorza:2013dav} and ALPS-II~\cite{Bahre:2013ywa}, and with the QCD axion parameter space as identified in~\cite{DiLuzio:2016sbl} (light green region). Note that in the pre-inflationary scenario, the predicted axion mass that accounts for all the dark matter has been computed to be $\sim 26 \mu$eV~\cite{Klaer:2017ond}, which lies precisely in the range where the prospects of this paper are more relevant. Our results illustrate that SKA could potentially improve upon existing helioscope constraints by roughly half an order of magnitude.

Before continuing, we comment briefly on the origin of the improvement in sensitivity relative to the original analysis performed in~\cite{Blout:2000uc}. At low masses, the contribution from stimulated emission is significant; specifically, considering $m_a \sim 10^{-6}$ eV, stimulated emission enhances the signal by roughly a factor of $10^3$ (which reduces to a factor $\sim 10$ for $m_a=10^{-4}$ eV). With regard to telescope sensitivity, SKA-Mid offers more than one order of magnitude reduction in system temperature and an enhancement in the effective area of $\sim 2$ orders of magnitude. Lastly, the discovery of new dSphs in recent years has provided more promising astrophysical targets, subsequently allowing for a 1-2 order of magnitude improvement of the overall observed flux. 

\section{Discussion}

In this letter, we have demonstrated that axion decay can be competitive with axion-photon conversion for axion masses probed by radio telescopes, depending on the astrophysical environment. Therefore, it should be included in the sensitivity studies searching for cold dark matter axions with radio surveys. Among the many candidate astrophysical sources, we have identified the best nearby dwarf spheroidal galaxies as a function of the telescope field of view.  We  have shown that SKA can identify axion cold dark matter by observing the axion decay inside nearby dwarf spheroidal galaxies. Our proposal complements new projects searching for cold dark matter axions in the unexplored mass region near $m_a \sim 10^{-4} $eV \cite{Irastorza:2018dyq}. Furthermore, we emphasize that the sensitivity obtained here is competitive with the projected coverage of ALPS-II, although subdominant to that of IAXO~\cite{Irastorza:2013dav, Bahre:2013ywa, Graham:2015ouw}. 

It should be understood that the projected sensitivity presented in this work represents what a particular \emph{planned} experiment can accomplished with \emph{known} astrophysical objects. In recent years, experiments such as the Dark Energy Survey have dramatically increased the rate of discovery of Milky Way dwarf galaxies, particularly those that offer promise for the indirect detection of dark matter~\cite{Koposov:2015cua,Bechtol:2015cbp,Drlica-Wagner:2015ufc}. With ongoing and future experiments such as Gaia~\cite{2018arXiv180501839M} and the Large Synoptic Survey Telescope~\cite{Ivezic:2008fe} providing unprecedented sensitivity to the gravitational effects of dark matter in the Galaxy, one may expect the rate of dwarf discovery to continue increasing. If more promising dwarf targets are observed in the near future, the projected sensitivity shown here may prove to be conservative.  It is important to bare in mind that future instruments intended to perform this search will require good frequency resolution, with a Nyquist width of the autocorrelation spectrometer of $\sim$ 2 MHz at 100 GHz. Finally, we emphasize that even in the absence of an axion detection, additional science can be obtained from observations of dSphs --  for example, molecular line searches in dSphs can enhance our current understanding of the limited star formation in these objects~\cite{0004-637X-850-1-54}.
  
%The sensitivity region presented here can improved in several ways. Already discovered dwarf spheroidal galaxies may not exhaust the existing ones. New dwarfs can be discovered in the exhaustive exploration of our galactic neighborhood by Gaia \cite{2018arXiv180501839M}. SKA and other future radio surveys have the potential to explore dwarf galaxies in wider frequency bands. In particular, future radio telescopes with sensitivity to the high frequency region can better probe the well motivated QCD axion parameter space. The instruments will require good frequency resolution, with Nyquist width of the autocorrelation spectrometer of $\sim$2 MHz at 100 GHz, and will be able to probe axion masses and couplings non accesible by other means nowadays. Needless to say that more science would come from pointing radiotelescopes to dwarf spheroidal galaxies, for example, searches for molecular lines in the gas clouds connected to star formation \cite{0004-637X-850-1-54}. 

\section{Acknowledgements}
CPG thanks initial discussions on Ref.~\cite{Blout:2000uc} with Matteo Viel and early work with Aaron Vincent and Francisco Villaescusa. The authors would also like to thank the following individuals for their useful comments and discussions: Javier Redondo, Jordi Miralda, Jorge Pe\~{n}arrubia, Leslie Rosenberg, Jose Carlos Guirado, Ivan Marti Vidal, Dan Hooper, Alfredo Urbano, Marco Taoso, Mauro Valli, Valentina De Romeri, Fernando Ballesteros, Marco Regis, Carlos Hernandez-Monteagudo, Jose Francisco Gomez and the SOM group at IFIC in Valencia. We were supported by PROMETEO II/2014/050 of Generalitat Valenciana, FPA2014-57816-P and FPA2017-85985-P of MINECO and by the European Union's Horizon 2020 research and innovation program under H2020-MSCA-ITN-2015//674896-ELUSIVES and H2020-MSCA-RISE-2015.

%SKA radiometers bandwidth 

%%%%%%%%%%%%%%%%%%%%%%%%%%%%%%%%%%%%%%%%%%%%%%%%%%%%%
\bibliography{biblio}

%merlin.mbs apsrev4-1.bst 2010-07-25 4.21a (PWD, AO, DPC) hacked
%Control: key (0)
%Control: author (8) initials jnrlst
%Control: editor formatted (1) identically to author
%Control: production of article title (-1) disabled
%Control: page (0) single
%Control: year (1) truncated
%Control: production of eprint (0) enabled
\begin{thebibliography}{59}%
\makeatletter
\providecommand \@ifxundefined [1]{%
 \@ifx{#1\undefined}
}%
\providecommand \@ifnum [1]{%
 \ifnum #1\expandafter \@firstoftwo
 \else \expandafter \@secondoftwo
 \fi
}%
\providecommand \@ifx [1]{%
 \ifx #1\expandafter \@firstoftwo
 \else \expandafter \@secondoftwo
 \fi
}%
\providecommand \natexlab [1]{#1}%
\providecommand \enquote  [1]{``#1''}%
\providecommand \bibnamefont  [1]{#1}%
\providecommand \bibfnamefont [1]{#1}%
\providecommand \citenamefont [1]{#1}%
\providecommand \href@noop [0]{\@secondoftwo}%
\providecommand \href [0]{\begingroup \@sanitize@url \@href}%
\providecommand \@href[1]{\@@startlink{#1}\@@href}%
\providecommand \@@href[1]{\endgroup#1\@@endlink}%
\providecommand \@sanitize@url [0]{\catcode `\\12\catcode `\$12\catcode
  `\&12\catcode `\#12\catcode `\^12\catcode `\_12\catcode `\%12\relax}%
\providecommand \@@startlink[1]{}%
\providecommand \@@endlink[0]{}%
\providecommand \url  [0]{\begingroup\@sanitize@url \@url }%
\providecommand \@url [1]{\endgroup\@href {#1}{\urlprefix }}%
\providecommand \urlprefix  [0]{URL }%
\providecommand \Eprint [0]{\href }%
\providecommand \doibase [0]{http://dx.doi.org/}%
\providecommand \selectlanguage [0]{\@gobble}%
\providecommand \bibinfo  [0]{\@secondoftwo}%
\providecommand \bibfield  [0]{\@secondoftwo}%
\providecommand \translation [1]{[#1]}%
\providecommand \BibitemOpen [0]{}%
\providecommand \bibitemStop [0]{}%
\providecommand \bibitemNoStop [0]{.\EOS\space}%
\providecommand \EOS [0]{\spacefactor3000\relax}%
\providecommand \BibitemShut  [1]{\csname bibitem#1\endcsname}%
\let\auto@bib@innerbib\@empty
%</preamble>
\bibitem [{\citenamefont {Peccei}\ and\ \citenamefont
  {Quinn}(1977{\natexlab{a}})}]{Peccei:1977hh}%
  \BibitemOpen
  \bibfield  {author} {\bibinfo {author} {\bibfnamefont {R.~D.}\ \bibnamefont
  {Peccei}}\ and\ \bibinfo {author} {\bibfnamefont {H.~R.}\ \bibnamefont
  {Quinn}},\ }\href {\doibase 10.1103/PhysRevLett.38.1440} {\bibfield
  {journal} {\bibinfo  {journal} {Phys. Rev. Lett.}\ }\textbf {\bibinfo
  {volume} {38}},\ \bibinfo {pages} {1440} (\bibinfo {year}
  {1977}{\natexlab{a}})},\ \bibinfo {note} {[,328(1977)]}\BibitemShut {NoStop}%
%%CITATION = PRLTA,38,1440;%%
\bibitem [{\citenamefont {Peccei}\ and\ \citenamefont
  {Quinn}(1977{\natexlab{b}})}]{Peccei:1977ur}%
  \BibitemOpen
  \bibfield  {author} {\bibinfo {author} {\bibfnamefont {R.~D.}\ \bibnamefont
  {Peccei}}\ and\ \bibinfo {author} {\bibfnamefont {H.~R.}\ \bibnamefont
  {Quinn}},\ }\href {\doibase 10.1103/PhysRevD.16.1791} {\bibfield  {journal}
  {\bibinfo  {journal} {Phys. Rev.}\ }\textbf {\bibinfo {volume} {D16}},\
  \bibinfo {pages} {1791} (\bibinfo {year} {1977}{\natexlab{b}})}\BibitemShut
  {NoStop}%
%%CITATION = PHRVA,D16,1791;%%
\bibitem [{\citenamefont {Weinberg}(1978)}]{Weinberg:1977ma}%
  \BibitemOpen
  \bibfield  {author} {\bibinfo {author} {\bibfnamefont {S.}~\bibnamefont
  {Weinberg}},\ }\href {\doibase 10.1103/PhysRevLett.40.223} {\bibfield
  {journal} {\bibinfo  {journal} {Phys. Rev. Lett.}\ }\textbf {\bibinfo
  {volume} {40}},\ \bibinfo {pages} {223} (\bibinfo {year} {1978})}\BibitemShut
  {NoStop}%
%%CITATION = PRLTA,40,223;%%
\bibitem [{\citenamefont {Wilczek}(1978)}]{Wilczek:1977pj}%
  \BibitemOpen
  \bibfield  {author} {\bibinfo {author} {\bibfnamefont {F.}~\bibnamefont
  {Wilczek}},\ }\href {\doibase 10.1103/PhysRevLett.40.279} {\bibfield
  {journal} {\bibinfo  {journal} {Phys. Rev. Lett.}\ }\textbf {\bibinfo
  {volume} {40}},\ \bibinfo {pages} {279} (\bibinfo {year} {1978})}\BibitemShut
  {NoStop}%
%%CITATION = PRLTA,40,279;%%
\bibitem [{\citenamefont {Peccei}(2008)}]{Peccei:2006as}%
  \BibitemOpen
  \bibfield  {author} {\bibinfo {author} {\bibfnamefont {R.~D.}\ \bibnamefont
  {Peccei}},\ }\bibfield  {booktitle} {\emph {\bibinfo {booktitle} {{Axions:
  Theory, cosmology, and experimental searches. Proceedings, 1st Joint
  ILIAS-CERN-CAST axion training, Geneva, Switzerland, November 30-December 2,
  2005}}},\ }\href {\doibase 10.1007/978-3-540-73518-2_1} {\bibfield  {journal}
  {\bibinfo  {journal} {Lect. Notes Phys.}\ }\textbf {\bibinfo {volume}
  {741}},\ \bibinfo {pages} {3} (\bibinfo {year} {2008})},\ \bibinfo {note}
  {[,3(2006)]},\ \Eprint {http://arxiv.org/abs/hep-ph/0607268}
  {arXiv:hep-ph/0607268 [hep-ph]} \BibitemShut {NoStop}%
%%CITATION = HEP-PH/0607268;%%
\bibitem [{\citenamefont {Preskill}\ \emph {et~al.}(1983)\citenamefont
  {Preskill}, \citenamefont {Wise},\ and\ \citenamefont
  {Wilczek}}]{Preskill:1982cy}%
  \BibitemOpen
  \bibfield  {author} {\bibinfo {author} {\bibfnamefont {J.}~\bibnamefont
  {Preskill}}, \bibinfo {author} {\bibfnamefont {M.~B.}\ \bibnamefont {Wise}},
  \ and\ \bibinfo {author} {\bibfnamefont {F.}~\bibnamefont {Wilczek}},\ }\href
  {\doibase 10.1016/0370-2693(83)90637-8} {\bibfield  {journal} {\bibinfo
  {journal} {Phys. Lett.}\ }\textbf {\bibinfo {volume} {B120}},\ \bibinfo
  {pages} {127} (\bibinfo {year} {1983})},\ \bibinfo {note}
  {[,URL(1982)]}\BibitemShut {NoStop}%
%%CITATION = PHLTA,B120,127;%%
\bibitem [{\citenamefont {Dine}\ and\ \citenamefont
  {Fischler}(1983)}]{Dine:1982ah}%
  \BibitemOpen
  \bibfield  {author} {\bibinfo {author} {\bibfnamefont {M.}~\bibnamefont
  {Dine}}\ and\ \bibinfo {author} {\bibfnamefont {W.}~\bibnamefont
  {Fischler}},\ }\href {\doibase 10.1016/0370-2693(83)90639-1} {\bibfield
  {journal} {\bibinfo  {journal} {Phys. Lett.}\ }\textbf {\bibinfo {volume}
  {B120}},\ \bibinfo {pages} {137} (\bibinfo {year} {1983})},\ \bibinfo {note}
  {[,URL(1982)]}\BibitemShut {NoStop}%
%%CITATION = PHLTA,B120,137;%%
\bibitem [{\citenamefont {Abbott}\ and\ \citenamefont
  {Sikivie}(1983)}]{Abbott:1982af}%
  \BibitemOpen
  \bibfield  {author} {\bibinfo {author} {\bibfnamefont {L.~F.}\ \bibnamefont
  {Abbott}}\ and\ \bibinfo {author} {\bibfnamefont {P.}~\bibnamefont
  {Sikivie}},\ }\href {\doibase 10.1016/0370-2693(83)90638-X} {\bibfield
  {journal} {\bibinfo  {journal} {Phys. Lett.}\ }\textbf {\bibinfo {volume}
  {B120}},\ \bibinfo {pages} {133} (\bibinfo {year} {1983})},\ \bibinfo {note}
  {[,URL(1982)]}\BibitemShut {NoStop}%
%%CITATION = PHLTA,B120,133;%%
\bibitem [{\citenamefont {Davis}(1986)}]{Davis:1986xc}%
  \BibitemOpen
  \bibfield  {author} {\bibinfo {author} {\bibfnamefont {R.~L.}\ \bibnamefont
  {Davis}},\ }\href {\doibase 10.1016/0370-2693(86)90300-X} {\bibfield
  {journal} {\bibinfo  {journal} {Phys. Lett.}\ }\textbf {\bibinfo {volume}
  {B180}},\ \bibinfo {pages} {225} (\bibinfo {year} {1986})}\BibitemShut
  {NoStop}%
%%CITATION = PHLTA,B180,225;%%
\bibitem [{\citenamefont {Lyth}(1992)}]{Lyth:1991bb}%
  \BibitemOpen
  \bibfield  {author} {\bibinfo {author} {\bibfnamefont {D.~H.}\ \bibnamefont
  {Lyth}},\ }\href {\doibase 10.1016/0370-2693(92)91590-6} {\bibfield
  {journal} {\bibinfo  {journal} {Phys. Lett.}\ }\textbf {\bibinfo {volume}
  {B275}},\ \bibinfo {pages} {279} (\bibinfo {year} {1992})}\BibitemShut
  {NoStop}%
%%CITATION = PHLTA,B275,279;%%
\bibitem [{\citenamefont {Irastorza}\ and\ \citenamefont
  {Redondo}(2018)}]{Irastorza:2018dyq}%
  \BibitemOpen
  \bibfield  {author} {\bibinfo {author} {\bibfnamefont {I.~G.}\ \bibnamefont
  {Irastorza}}\ and\ \bibinfo {author} {\bibfnamefont {J.}~\bibnamefont
  {Redondo}},\ }\href@noop {} {\  (\bibinfo {year} {2018})},\ \Eprint
  {http://arxiv.org/abs/1801.08127} {arXiv:1801.08127 [hep-ph]} \BibitemShut
  {NoStop}%
%%CITATION = ARXIV:1801.08127;%%
\bibitem [{\citenamefont {Kelley}\ and\ \citenamefont
  {Quinn}(2017)}]{Kelley:2017vaa}%
  \BibitemOpen
  \bibfield  {author} {\bibinfo {author} {\bibfnamefont {K.}~\bibnamefont
  {Kelley}}\ and\ \bibinfo {author} {\bibfnamefont {P.~J.}\ \bibnamefont
  {Quinn}},\ }\href {\doibase 10.3847/2041-8213/aa808d} {\bibfield  {journal}
  {\bibinfo  {journal} {Astrophys. J.}\ }\textbf {\bibinfo {volume} {845}},\
  \bibinfo {pages} {L4} (\bibinfo {year} {2017})},\ \Eprint
  {http://arxiv.org/abs/1708.01399} {arXiv:1708.01399 [astro-ph.CO]}
  \BibitemShut {NoStop}%
%%CITATION = ARXIV:1708.01399;%%
\bibitem [{\citenamefont {Huang}\ \emph {et~al.}(2018)\citenamefont {Huang},
  \citenamefont {Kadota}, \citenamefont {Sekiguchi},\ and\ \citenamefont
  {Tashiro}}]{Huang:2018lxq}%
  \BibitemOpen
  \bibfield  {author} {\bibinfo {author} {\bibfnamefont {F.~P.}\ \bibnamefont
  {Huang}}, \bibinfo {author} {\bibfnamefont {K.}~\bibnamefont {Kadota}},
  \bibinfo {author} {\bibfnamefont {T.}~\bibnamefont {Sekiguchi}}, \ and\
  \bibinfo {author} {\bibfnamefont {H.}~\bibnamefont {Tashiro}},\ }\href@noop
  {} {\  (\bibinfo {year} {2018})},\ \Eprint {http://arxiv.org/abs/1803.08230}
  {arXiv:1803.08230 [hep-ph]} \BibitemShut {NoStop}%
%%CITATION = ARXIV:1803.08230;%%
\bibitem [{\citenamefont {Hook}\ \emph {et~al.}(2018)\citenamefont {Hook},
  \citenamefont {Kahn}, \citenamefont {Safdi},\ and\ \citenamefont
  {Sun}}]{Hook:2018iia}%
  \BibitemOpen
  \bibfield  {author} {\bibinfo {author} {\bibfnamefont {A.}~\bibnamefont
  {Hook}}, \bibinfo {author} {\bibfnamefont {Y.}~\bibnamefont {Kahn}}, \bibinfo
  {author} {\bibfnamefont {B.~R.}\ \bibnamefont {Safdi}}, \ and\ \bibinfo
  {author} {\bibfnamefont {Z.}~\bibnamefont {Sun}},\ }\href@noop {} {\
  (\bibinfo {year} {2018})},\ \Eprint {http://arxiv.org/abs/1804.03145}
  {arXiv:1804.03145 [hep-ph]} \BibitemShut {NoStop}%
%%CITATION = ARXIV:1804.03145;%%
\bibitem [{\citenamefont {Blout}\ \emph {et~al.}(2001)\citenamefont {Blout},
  \citenamefont {Daw}, \citenamefont {Decowski}, \citenamefont {Ho},
  \citenamefont {Rosenberg},\ and\ \citenamefont {Yu}}]{Blout:2000uc}%
  \BibitemOpen
  \bibfield  {author} {\bibinfo {author} {\bibfnamefont {B.~D.}\ \bibnamefont
  {Blout}}, \bibinfo {author} {\bibfnamefont {E.~J.}\ \bibnamefont {Daw}},
  \bibinfo {author} {\bibfnamefont {M.~P.}\ \bibnamefont {Decowski}}, \bibinfo
  {author} {\bibfnamefont {P.~T.~P.}\ \bibnamefont {Ho}}, \bibinfo {author}
  {\bibfnamefont {L.~J.}\ \bibnamefont {Rosenberg}}, \ and\ \bibinfo {author}
  {\bibfnamefont {D.~B.}\ \bibnamefont {Yu}},\ }\href {\doibase 10.1086/318310}
  {\bibfield  {journal} {\bibinfo  {journal} {Astrophys. J.}\ }\textbf
  {\bibinfo {volume} {546}},\ \bibinfo {pages} {825} (\bibinfo {year}
  {2001})},\ \Eprint {http://arxiv.org/abs/astro-ph/0006310}
  {arXiv:astro-ph/0006310 [astro-ph]} \BibitemShut {NoStop}%
%%CITATION = ASTRO-PH/0006310;%%
\bibitem [{Win()}]{WinNT}%
  \BibitemOpen
  \href@noop {} {\enquote {\bibinfo {title} {{SKA Whitepaper}, howpublished =
  {\url{https://www.skatelescope.org/wp-content/uploads/2014/03/ska-tel-sko-0000308_ska1_system_baseline_v2_descriptionrev01-part-1-signed.pdf}},
  note = {Accessed: 2018-04-06}},}\ }\BibitemShut {NoStop}%
\bibitem [{\citenamefont {Shokair}\ \emph {et~al.}(2014)\citenamefont {Shokair}
  \emph {et~al.}}]{Shokair:2014rna}%
  \BibitemOpen
  \bibfield  {author} {\bibinfo {author} {\bibfnamefont {T.~M.}\ \bibnamefont
  {Shokair}} \emph {et~al.},\ }\href {\doibase 10.1142/S0217751X14430040}
  {\bibfield  {journal} {\bibinfo  {journal} {Int. J. Mod. Phys.}\ }\textbf
  {\bibinfo {volume} {A29}},\ \bibinfo {pages} {1443004} (\bibinfo {year}
  {2014})},\ \Eprint {http://arxiv.org/abs/1405.3685} {arXiv:1405.3685
  [physics.ins-det]} \BibitemShut {NoStop}%
%%CITATION = ARXIV:1405.3685;%%
\bibitem [{\citenamefont {Budker}\ \emph {et~al.}(2014)\citenamefont {Budker},
  \citenamefont {Graham}, \citenamefont {Ledbetter}, \citenamefont
  {Rajendran},\ and\ \citenamefont {Sushkov}}]{Budker:2013hfa}%
  \BibitemOpen
  \bibfield  {author} {\bibinfo {author} {\bibfnamefont {D.}~\bibnamefont
  {Budker}}, \bibinfo {author} {\bibfnamefont {P.~W.}\ \bibnamefont {Graham}},
  \bibinfo {author} {\bibfnamefont {M.}~\bibnamefont {Ledbetter}}, \bibinfo
  {author} {\bibfnamefont {S.}~\bibnamefont {Rajendran}}, \ and\ \bibinfo
  {author} {\bibfnamefont {A.}~\bibnamefont {Sushkov}},\ }\href {\doibase
  10.1103/PhysRevX.4.021030} {\bibfield  {journal} {\bibinfo  {journal} {Phys.
  Rev.}\ }\textbf {\bibinfo {volume} {X4}},\ \bibinfo {pages} {021030}
  (\bibinfo {year} {2014})},\ \Eprint {http://arxiv.org/abs/1306.6089}
  {arXiv:1306.6089 [hep-ph]} \BibitemShut {NoStop}%
%%CITATION = ARXIV:1306.6089;%%
\bibitem [{\citenamefont {Brubaker}\ \emph
  {et~al.}(2017{\natexlab{a}})\citenamefont {Brubaker} \emph
  {et~al.}}]{Brubaker:2016ktl}%
  \BibitemOpen
  \bibfield  {author} {\bibinfo {author} {\bibfnamefont {B.~M.}\ \bibnamefont
  {Brubaker}} \emph {et~al.},\ }\href {\doibase 10.1103/PhysRevLett.118.061302}
  {\bibfield  {journal} {\bibinfo  {journal} {Phys. Rev. Lett.}\ }\textbf
  {\bibinfo {volume} {118}},\ \bibinfo {pages} {061302} (\bibinfo {year}
  {2017}{\natexlab{a}})},\ \Eprint {http://arxiv.org/abs/1610.02580}
  {arXiv:1610.02580 [astro-ph.CO]} \BibitemShut {NoStop}%
%%CITATION = ARXIV:1610.02580;%%
\bibitem [{\citenamefont {Al~Kenany}\ \emph {et~al.}(2017)\citenamefont
  {Al~Kenany} \emph {et~al.}}]{Kenany:2016tta}%
  \BibitemOpen
  \bibfield  {author} {\bibinfo {author} {\bibfnamefont {S.}~\bibnamefont
  {Al~Kenany}} \emph {et~al.},\ }\href {\doibase 10.1016/j.nima.2017.02.012}
  {\bibfield  {journal} {\bibinfo  {journal} {Nucl. Instrum. Meth.}\ }\textbf
  {\bibinfo {volume} {A854}},\ \bibinfo {pages} {11} (\bibinfo {year}
  {2017})},\ \Eprint {http://arxiv.org/abs/1611.07123} {arXiv:1611.07123
  [physics.ins-det]} \BibitemShut {NoStop}%
%%CITATION = ARXIV:1611.07123;%%
\bibitem [{\citenamefont {Majorovits}\ and\ \citenamefont
  {Redondo}(2017)}]{Majorovits:2016yvk}%
  \BibitemOpen
  \bibfield  {author} {\bibinfo {author} {\bibfnamefont {B.}~\bibnamefont
  {Majorovits}}\ and\ \bibinfo {author} {\bibfnamefont {J.}~\bibnamefont
  {Redondo}} (\bibinfo {collaboration} {MADMAX Working Group}),\ }in\ \href
  {\doibase 10.3204/DESY-PROC-2009-03/Majorovits_Bela} {\emph {\bibinfo
  {booktitle} {{Proceedings, 12th Patras Workshop on Axions, WIMPs and WISPs
  (PATRAS 2016): Jeju Island, South Korea, June 20-24, 2016}}}}\ (\bibinfo
  {year} {2017})\ pp.\ \bibinfo {pages} {94--97},\ \Eprint
  {http://arxiv.org/abs/1611.04549} {arXiv:1611.04549 [astro-ph.IM]}
  \BibitemShut {NoStop}%
%%CITATION = ARXIV:1611.04549;%%
\bibitem [{\citenamefont {Brubaker}\ \emph
  {et~al.}(2017{\natexlab{b}})\citenamefont {Brubaker}, \citenamefont {Zhong},
  \citenamefont {Lamoreaux}, \citenamefont {Lehnert},\ and\ \citenamefont {van
  Bibber}}]{Brubaker:2017rna}%
  \BibitemOpen
  \bibfield  {author} {\bibinfo {author} {\bibfnamefont {B.~M.}\ \bibnamefont
  {Brubaker}}, \bibinfo {author} {\bibfnamefont {L.}~\bibnamefont {Zhong}},
  \bibinfo {author} {\bibfnamefont {S.~K.}\ \bibnamefont {Lamoreaux}}, \bibinfo
  {author} {\bibfnamefont {K.~W.}\ \bibnamefont {Lehnert}}, \ and\ \bibinfo
  {author} {\bibfnamefont {K.~A.}\ \bibnamefont {van Bibber}},\ }\href
  {\doibase 10.1103/PhysRevD.96.123008} {\bibfield  {journal} {\bibinfo
  {journal} {Phys. Rev.}\ }\textbf {\bibinfo {volume} {D96}},\ \bibinfo {pages}
  {123008} (\bibinfo {year} {2017}{\natexlab{b}})},\ \Eprint
  {http://arxiv.org/abs/1706.08388} {arXiv:1706.08388 [astro-ph.IM]}
  \BibitemShut {NoStop}%
%%CITATION = ARXIV:1706.08388;%%
\bibitem [{\citenamefont {Sigl}(2017)}]{Sigl:2017sew}%
  \BibitemOpen
  \bibfield  {author} {\bibinfo {author} {\bibfnamefont {G.}~\bibnamefont
  {Sigl}},\ }\href {\doibase 10.1103/PhysRevD.96.103014} {\bibfield  {journal}
  {\bibinfo  {journal} {Phys. Rev.}\ }\textbf {\bibinfo {volume} {D96}},\
  \bibinfo {pages} {103014} (\bibinfo {year} {2017})},\ \Eprint
  {http://arxiv.org/abs/1708.08908} {arXiv:1708.08908 [astro-ph.HE]}
  \BibitemShut {NoStop}%
%%CITATION = ARXIV:1708.08908;%%
\bibitem [{\citenamefont {Sikivie}(1983)}]{Sikivie:1983ip}%
  \BibitemOpen
  \bibfield  {author} {\bibinfo {author} {\bibfnamefont {P.}~\bibnamefont
  {Sikivie}},\ }\bibfield  {booktitle} {\emph {\bibinfo {booktitle} {{Particle
  physics and cosmology: Dark matter}}},\ }\href {\doibase
  10.1103/PhysRevLett.51.1415, 10.1103/PhysRevLett.52.695.2} {\bibfield
  {journal} {\bibinfo  {journal} {Phys. Rev. Lett.}\ }\textbf {\bibinfo
  {volume} {51}},\ \bibinfo {pages} {1415} (\bibinfo {year} {1983})},\ \bibinfo
  {note} {[,321(1983)]}\BibitemShut {NoStop}%
%%CITATION = PRLTA,51,1415;%%
\bibitem [{\citenamefont {LaRosa}\ \emph {et~al.}(2006)\citenamefont {LaRosa},
  \citenamefont {Shore}, \citenamefont {Joseph}, \citenamefont {Lazio},\ and\
  \citenamefont {Kassim}}]{LaRosa2006}%
  \BibitemOpen
  \bibfield  {author} {\bibinfo {author} {\bibfnamefont {T.~N.}\ \bibnamefont
  {LaRosa}}, \bibinfo {author} {\bibfnamefont {S.~N.}\ \bibnamefont {Shore}},
  \bibinfo {author} {\bibfnamefont {T.}~\bibnamefont {Joseph}}, \bibinfo
  {author} {\bibfnamefont {W.}~\bibnamefont {Lazio}}, \ and\ \bibinfo {author}
  {\bibfnamefont {N.~E.}\ \bibnamefont {Kassim}},\ }\href
  {http://stacks.iop.org/1742-6596/54/i=1/a=002} {\bibfield  {journal}
  {\bibinfo  {journal} {Journal of Physics: Conference Series}\ }\textbf
  {\bibinfo {volume} {54}},\ \bibinfo {pages} {10} (\bibinfo {year}
  {2006})}\BibitemShut {NoStop}%
\bibitem [{\citenamefont {Beck}(2009)}]{Beck:2008ty}%
  \BibitemOpen
  \bibfield  {author} {\bibinfo {author} {\bibfnamefont {R.}~\bibnamefont
  {Beck}},\ }\bibfield  {booktitle} {\emph {\bibinfo {booktitle} {{Proceedings,
  4th Heidelberg International Symposium on High-Energy Gamma Ray Astronomy:
  Heidelberg, Germany, July 7-11, 2008}}},\ }\href {\doibase 10.1063/1.3076806}
  {\bibfield  {journal} {\bibinfo  {journal} {AIP Conf. Proc.}\ }\textbf
  {\bibinfo {volume} {1085}},\ \bibinfo {pages} {83} (\bibinfo {year}
  {2009})},\ \Eprint {http://arxiv.org/abs/0810.2923} {arXiv:0810.2923
  [astro-ph]} \BibitemShut {NoStop}%
%%CITATION = ARXIV:0810.2923;%%
\bibitem [{\citenamefont {Ferriere}(2009)}]{Ferriere:2009dh}%
  \BibitemOpen
  \bibfield  {author} {\bibinfo {author} {\bibfnamefont {K.}~\bibnamefont
  {Ferriere}},\ }\href {\doibase 10.1051/0004-6361/200912617} {\bibfield
  {journal} {\bibinfo  {journal} {Astron. Astrophys.}\ }\textbf {\bibinfo
  {volume} {505}},\ \bibinfo {pages} {1183} (\bibinfo {year} {2009})},\ \Eprint
  {http://arxiv.org/abs/0908.2037} {arXiv:0908.2037 [astro-ph.GA]} \BibitemShut
  {NoStop}%
%%CITATION = ARXIV:0908.2037;%%
\bibitem [{\citenamefont {Han}(2009)}]{Han:2009ts}%
  \BibitemOpen
  \bibfield  {author} {\bibinfo {author} {\bibfnamefont {J.~L.}\ \bibnamefont
  {Han}},\ }\bibfield  {booktitle} {\emph {\bibinfo {booktitle} {{Proceedings,
  IAU Symposium 259: Cosmic Magnetic Fields: From Planets, to Stars and
  Galaxies: Puerto Santiago, Tenerife, Spain, November 3-7, 2008}}},\ }\href
  {\doibase 10.1017/S1743921309031123} {\bibfield  {journal} {\bibinfo
  {journal} {IAU Symp.}\ }\textbf {\bibinfo {volume} {259}},\ \bibinfo {pages}
  {455} (\bibinfo {year} {2009})},\ \Eprint {http://arxiv.org/abs/0901.1165}
  {arXiv:0901.1165 [astro-ph.GA]} \BibitemShut {NoStop}%
%%CITATION = ARXIV:0901.1165;%%
\bibitem [{\citenamefont {Beck}\ and\ \citenamefont
  {Wielebinski}(2013)}]{Beck:2013bxa}%
  \BibitemOpen
  \bibfield  {author} {\bibinfo {author} {\bibfnamefont {R.}~\bibnamefont
  {Beck}}\ and\ \bibinfo {author} {\bibfnamefont {R.}~\bibnamefont
  {Wielebinski}}\ }(\bibinfo {year} {2013})\ pp.\ \bibinfo {pages} {641--723},\
  \Eprint {http://arxiv.org/abs/1302.5663} {arXiv:1302.5663 [astro-ph.GA]}
  \BibitemShut {NoStop}%
%%CITATION = ARXIV:1302.5663;%%
\bibitem [{\citenamefont {Regis}\ \emph {et~al.}(2015)\citenamefont {Regis},
  \citenamefont {Richter}, \citenamefont {Colafrancesco}, \citenamefont
  {Profumo}, \citenamefont {de~Blok},\ and\ \citenamefont
  {Massardi}}]{Regis:2014koa}%
  \BibitemOpen
  \bibfield  {author} {\bibinfo {author} {\bibfnamefont {M.}~\bibnamefont
  {Regis}}, \bibinfo {author} {\bibfnamefont {L.}~\bibnamefont {Richter}},
  \bibinfo {author} {\bibfnamefont {S.}~\bibnamefont {Colafrancesco}}, \bibinfo
  {author} {\bibfnamefont {S.}~\bibnamefont {Profumo}}, \bibinfo {author}
  {\bibfnamefont {W.~J.~G.}\ \bibnamefont {de~Blok}}, \ and\ \bibinfo {author}
  {\bibfnamefont {M.}~\bibnamefont {Massardi}},\ }\href {\doibase
  10.1093/mnras/stv127} {\bibfield  {journal} {\bibinfo  {journal} {Mon. Not.
  Roy. Astron. Soc.}\ }\textbf {\bibinfo {volume} {448}},\ \bibinfo {pages}
  {3747} (\bibinfo {year} {2015})},\ \Eprint {http://arxiv.org/abs/1407.5482}
  {arXiv:1407.5482 [astro-ph.GA]} \BibitemShut {NoStop}%
%%CITATION = ARXIV:1407.5482;%%
\bibitem [{\citenamefont {Arlen}\ \emph {et~al.}(2014)\citenamefont {Arlen},
  \citenamefont {Vassiliev}, \citenamefont {Weisgarber}, \citenamefont
  {Wakely},\ and\ \citenamefont {Shafi}}]{Arlen:2012iy}%
  \BibitemOpen
  \bibfield  {author} {\bibinfo {author} {\bibfnamefont {T.~C.}\ \bibnamefont
  {Arlen}}, \bibinfo {author} {\bibfnamefont {V.~V.}\ \bibnamefont
  {Vassiliev}}, \bibinfo {author} {\bibfnamefont {T.}~\bibnamefont
  {Weisgarber}}, \bibinfo {author} {\bibfnamefont {S.~P.}\ \bibnamefont
  {Wakely}}, \ and\ \bibinfo {author} {\bibfnamefont {S.~Y.}\ \bibnamefont
  {Shafi}},\ }\href {\doibase 10.1088/0004-637X/796/1/18} {\bibfield  {journal}
  {\bibinfo  {journal} {Astrophys. J.}\ }\textbf {\bibinfo {volume} {796}},\
  \bibinfo {pages} {18} (\bibinfo {year} {2014})},\ \Eprint
  {http://arxiv.org/abs/1210.2802} {arXiv:1210.2802 [astro-ph.HE]} \BibitemShut
  {NoStop}%
%%CITATION = ARXIV:1210.2802;%%
\bibitem [{\citenamefont {Finke}\ \emph {et~al.}(2015)\citenamefont {Finke},
  \citenamefont {Reyes}, \citenamefont {Georganopoulos}, \citenamefont
  {Reynolds}, \citenamefont {Ajello}, \citenamefont {Fegan},\ and\
  \citenamefont {McCann}}]{Finke:2015ona}%
  \BibitemOpen
  \bibfield  {author} {\bibinfo {author} {\bibfnamefont {J.~D.}\ \bibnamefont
  {Finke}}, \bibinfo {author} {\bibfnamefont {L.~C.}\ \bibnamefont {Reyes}},
  \bibinfo {author} {\bibfnamefont {M.}~\bibnamefont {Georganopoulos}},
  \bibinfo {author} {\bibfnamefont {K.}~\bibnamefont {Reynolds}}, \bibinfo
  {author} {\bibfnamefont {M.}~\bibnamefont {Ajello}}, \bibinfo {author}
  {\bibfnamefont {S.~J.}\ \bibnamefont {Fegan}}, \ and\ \bibinfo {author}
  {\bibfnamefont {K.}~\bibnamefont {McCann}},\ }\href {\doibase
  10.1088/0004-637X/814/1/20} {\bibfield  {journal} {\bibinfo  {journal}
  {Astrophys. J.}\ }\textbf {\bibinfo {volume} {814}},\ \bibinfo {pages} {20}
  (\bibinfo {year} {2015})},\ \Eprint {http://arxiv.org/abs/1510.02485}
  {arXiv:1510.02485 [astro-ph.HE]} \BibitemShut {NoStop}%
%%CITATION = ARXIV:1510.02485;%%
\bibitem [{SKA()}]{SKABase}%
  \BibitemOpen
  \href@noop {} {\enquote {\bibinfo {title} {{SKA Baseline Design},
  howpublished =
  {\url{https://www.skatelescope.org/wp-content/uploads/2013/08/ska-tel-sko-dd-001-1_baselinedesign1.pdf}},
  note = {Accessed: 2018-06-09}},}\ }\BibitemShut {NoStop}%
\bibitem [{\citenamefont {Geringer-Sameth}\ \emph {et~al.}(2015)\citenamefont
  {Geringer-Sameth}, \citenamefont {Koushiappas},\ and\ \citenamefont
  {Walker}}]{Geringer-Sameth:2014yza}%
  \BibitemOpen
  \bibfield  {author} {\bibinfo {author} {\bibfnamefont {A.}~\bibnamefont
  {Geringer-Sameth}}, \bibinfo {author} {\bibfnamefont {S.~M.}\ \bibnamefont
  {Koushiappas}}, \ and\ \bibinfo {author} {\bibfnamefont {M.}~\bibnamefont
  {Walker}},\ }\href {\doibase 10.1088/0004-637X/801/2/74} {\bibfield
  {journal} {\bibinfo  {journal} {Astrophys. J.}\ }\textbf {\bibinfo {volume}
  {801}},\ \bibinfo {pages} {74} (\bibinfo {year} {2015})},\ \Eprint
  {http://arxiv.org/abs/1408.0002} {arXiv:1408.0002 [astro-ph.CO]} \BibitemShut
  {NoStop}%
%%CITATION = ARXIV:1408.0002;%%
\bibitem [{\citenamefont {Bonnivard}\ \emph
  {et~al.}(2015{\natexlab{a}})\citenamefont {Bonnivard} \emph
  {et~al.}}]{Bonnivard:2015xpq}%
  \BibitemOpen
  \bibfield  {author} {\bibinfo {author} {\bibfnamefont {V.}~\bibnamefont
  {Bonnivard}} \emph {et~al.},\ }\href {\doibase 10.1093/mnras/stv1601}
  {\bibfield  {journal} {\bibinfo  {journal} {Mon. Not. Roy. Astron. Soc.}\
  }\textbf {\bibinfo {volume} {453}},\ \bibinfo {pages} {849} (\bibinfo {year}
  {2015}{\natexlab{a}})},\ \Eprint {http://arxiv.org/abs/1504.02048}
  {arXiv:1504.02048 [astro-ph.HE]} \BibitemShut {NoStop}%
%%CITATION = ARXIV:1504.02048;%%
\bibitem [{\citenamefont {Bonnivard}\ \emph
  {et~al.}(2015{\natexlab{b}})\citenamefont {Bonnivard}, \citenamefont
  {Combet}, \citenamefont {Maurin}, \citenamefont {Geringer-Sameth},
  \citenamefont {Koushiappas}, \citenamefont {Walker}, \citenamefont {Mateo},
  \citenamefont {Olszewski},\ and\ \citenamefont
  {Bailey~III}}]{Bonnivard:2015tta}%
  \BibitemOpen
  \bibfield  {author} {\bibinfo {author} {\bibfnamefont {V.}~\bibnamefont
  {Bonnivard}}, \bibinfo {author} {\bibfnamefont {C.}~\bibnamefont {Combet}},
  \bibinfo {author} {\bibfnamefont {D.}~\bibnamefont {Maurin}}, \bibinfo
  {author} {\bibfnamefont {A.}~\bibnamefont {Geringer-Sameth}}, \bibinfo
  {author} {\bibfnamefont {S.~M.}\ \bibnamefont {Koushiappas}}, \bibinfo
  {author} {\bibfnamefont {M.~G.}\ \bibnamefont {Walker}}, \bibinfo {author}
  {\bibfnamefont {M.}~\bibnamefont {Mateo}}, \bibinfo {author} {\bibfnamefont
  {E.~W.}\ \bibnamefont {Olszewski}}, \ and\ \bibinfo {author} {\bibfnamefont
  {J.~I.}\ \bibnamefont {Bailey~III}},\ }\href {\doibase
  10.1088/2041-8205/808/2/L36} {\bibfield  {journal} {\bibinfo  {journal}
  {Astrophys. J.}\ }\textbf {\bibinfo {volume} {808}},\ \bibinfo {pages} {L36}
  (\bibinfo {year} {2015}{\natexlab{b}})},\ \Eprint
  {http://arxiv.org/abs/1504.03309} {arXiv:1504.03309 [astro-ph.HE]}
  \BibitemShut {NoStop}%
%%CITATION = ARXIV:1504.03309;%%
\bibitem [{\citenamefont {Hayashi}\ \emph {et~al.}(2016)\citenamefont
  {Hayashi}, \citenamefont {Ichikawa}, \citenamefont {Matsumoto}, \citenamefont
  {Ibe}, \citenamefont {Ishigaki},\ and\ \citenamefont
  {Sugai}}]{Hayashi:2016kcy}%
  \BibitemOpen
  \bibfield  {author} {\bibinfo {author} {\bibfnamefont {K.}~\bibnamefont
  {Hayashi}}, \bibinfo {author} {\bibfnamefont {K.}~\bibnamefont {Ichikawa}},
  \bibinfo {author} {\bibfnamefont {S.}~\bibnamefont {Matsumoto}}, \bibinfo
  {author} {\bibfnamefont {M.}~\bibnamefont {Ibe}}, \bibinfo {author}
  {\bibfnamefont {M.~N.}\ \bibnamefont {Ishigaki}}, \ and\ \bibinfo {author}
  {\bibfnamefont {H.}~\bibnamefont {Sugai}},\ }\href {\doibase
  10.1093/mnras/stw1457} {\bibfield  {journal} {\bibinfo  {journal} {Mon. Not.
  Roy. Astron. Soc.}\ }\textbf {\bibinfo {volume} {461}},\ \bibinfo {pages}
  {2914} (\bibinfo {year} {2016})},\ \Eprint {http://arxiv.org/abs/1603.08046}
  {arXiv:1603.08046 [astro-ph.GA]} \BibitemShut {NoStop}%
%%CITATION = ARXIV:1603.08046;%%
\bibitem [{\citenamefont {Sanders}\ \emph {et~al.}(2016)\citenamefont
  {Sanders}, \citenamefont {Evans}, \citenamefont {Geringer-Sameth},\ and\
  \citenamefont {Dehnen}}]{Sanders:2016eie}%
  \BibitemOpen
  \bibfield  {author} {\bibinfo {author} {\bibfnamefont {J.~L.}\ \bibnamefont
  {Sanders}}, \bibinfo {author} {\bibfnamefont {N.~W.}\ \bibnamefont {Evans}},
  \bibinfo {author} {\bibfnamefont {A.}~\bibnamefont {Geringer-Sameth}}, \ and\
  \bibinfo {author} {\bibfnamefont {W.}~\bibnamefont {Dehnen}},\ }\href
  {\doibase 10.1103/PhysRevD.94.063521} {\bibfield  {journal} {\bibinfo
  {journal} {Phys. Rev.}\ }\textbf {\bibinfo {volume} {D94}},\ \bibinfo {pages}
  {063521} (\bibinfo {year} {2016})},\ \Eprint
  {http://arxiv.org/abs/1604.05493} {arXiv:1604.05493 [astro-ph.GA]}
  \BibitemShut {NoStop}%
%%CITATION = ARXIV:1604.05493;%%
\bibitem [{\citenamefont {Evans}\ \emph {et~al.}(2016)\citenamefont {Evans},
  \citenamefont {Sanders},\ and\ \citenamefont
  {Geringer-Sameth}}]{Evans:2016xwx}%
  \BibitemOpen
  \bibfield  {author} {\bibinfo {author} {\bibfnamefont {N.~W.}\ \bibnamefont
  {Evans}}, \bibinfo {author} {\bibfnamefont {J.~L.}\ \bibnamefont {Sanders}},
  \ and\ \bibinfo {author} {\bibfnamefont {A.}~\bibnamefont
  {Geringer-Sameth}},\ }\href {\doibase 10.1103/PhysRevD.93.103512} {\bibfield
  {journal} {\bibinfo  {journal} {Phys. Rev.}\ }\textbf {\bibinfo {volume}
  {D93}},\ \bibinfo {pages} {103512} (\bibinfo {year} {2016})},\ \Eprint
  {http://arxiv.org/abs/1604.05599} {arXiv:1604.05599 [astro-ph.GA]}
  \BibitemShut {NoStop}%
%%CITATION = ARXIV:1604.05599;%%
\bibitem [{\citenamefont {Hayashi}\ \emph {et~al.}(2018)\citenamefont
  {Hayashi}, \citenamefont {Fabrizio}, \citenamefont {Łokas}, \citenamefont
  {Bono}, \citenamefont {Monelli}, \citenamefont {Dall'Ora},\ and\
  \citenamefont {Stetson}}]{Hayashi:2018uop}%
  \BibitemOpen
  \bibfield  {author} {\bibinfo {author} {\bibfnamefont {K.}~\bibnamefont
  {Hayashi}}, \bibinfo {author} {\bibfnamefont {M.}~\bibnamefont {Fabrizio}},
  \bibinfo {author} {\bibfnamefont {E.~L.}\ \bibnamefont {Łokas}}, \bibinfo
  {author} {\bibfnamefont {G.}~\bibnamefont {Bono}}, \bibinfo {author}
  {\bibfnamefont {M.}~\bibnamefont {Monelli}}, \bibinfo {author} {\bibfnamefont
  {M.}~\bibnamefont {Dall'Ora}}, \ and\ \bibinfo {author} {\bibfnamefont
  {P.~B.}\ \bibnamefont {Stetson}},\ }\href@noop {} {\  (\bibinfo {year}
  {2018})},\ \Eprint {http://arxiv.org/abs/1804.01739} {arXiv:1804.01739
  [astro-ph.GA]} \BibitemShut {NoStop}%
%%CITATION = ARXIV:1804.01739;%%
\bibitem [{\citenamefont {Petac}\ \emph {et~al.}(2018)\citenamefont {Petac},
  \citenamefont {Ullio},\ and\ \citenamefont {Valli}}]{Petac:2018gue}%
  \BibitemOpen
  \bibfield  {author} {\bibinfo {author} {\bibfnamefont {M.}~\bibnamefont
  {Petac}}, \bibinfo {author} {\bibfnamefont {P.}~\bibnamefont {Ullio}}, \ and\
  \bibinfo {author} {\bibfnamefont {M.}~\bibnamefont {Valli}},\ }\href@noop {}
  {\  (\bibinfo {year} {2018})},\ \Eprint {http://arxiv.org/abs/1804.05052}
  {arXiv:1804.05052 [astro-ph.GA]} \BibitemShut {NoStop}%
%%CITATION = ARXIV:1804.05052;%%
\bibitem [{\citenamefont {Di~Luzio}\ \emph {et~al.}(2017)\citenamefont
  {Di~Luzio}, \citenamefont {Mescia},\ and\ \citenamefont
  {Nardi}}]{DiLuzio:2016sbl}%
  \BibitemOpen
  \bibfield  {author} {\bibinfo {author} {\bibfnamefont {L.}~\bibnamefont
  {Di~Luzio}}, \bibinfo {author} {\bibfnamefont {F.}~\bibnamefont {Mescia}}, \
  and\ \bibinfo {author} {\bibfnamefont {E.}~\bibnamefont {Nardi}},\ }\href
  {\doibase 10.1103/PhysRevLett.118.031801} {\bibfield  {journal} {\bibinfo
  {journal} {Phys. Rev. Lett.}\ }\textbf {\bibinfo {volume} {118}},\ \bibinfo
  {pages} {031801} (\bibinfo {year} {2017})},\ \Eprint
  {http://arxiv.org/abs/1610.07593} {arXiv:1610.07593 [hep-ph]} \BibitemShut
  {NoStop}%
%%CITATION = ARXIV:1610.07593;%%
\bibitem [{\citenamefont {Anastassopoulos}\ \emph {et~al.}(2017)\citenamefont
  {Anastassopoulos} \emph {et~al.}}]{Anastassopoulos:2017ftl}%
  \BibitemOpen
  \bibfield  {author} {\bibinfo {author} {\bibfnamefont {V.}~\bibnamefont
  {Anastassopoulos}} \emph {et~al.} (\bibinfo {collaboration} {CAST}),\ }\href
  {\doibase 10.1038/nphys4109} {\bibfield  {journal} {\bibinfo  {journal}
  {Nature Phys.}\ }\textbf {\bibinfo {volume} {13}},\ \bibinfo {pages} {584}
  (\bibinfo {year} {2017})},\ \Eprint {http://arxiv.org/abs/1705.02290}
  {arXiv:1705.02290 [hep-ex]} \BibitemShut {NoStop}%
%%CITATION = ARXIV:1705.02290;%%
\bibitem [{\citenamefont {Hagmann}\ \emph {et~al.}(1990)\citenamefont
  {Hagmann}, \citenamefont {Sikivie}, \citenamefont {Sullivan},\ and\
  \citenamefont {Tanner}}]{Hagmann:1990tj}%
  \BibitemOpen
  \bibfield  {author} {\bibinfo {author} {\bibfnamefont {C.}~\bibnamefont
  {Hagmann}}, \bibinfo {author} {\bibfnamefont {P.}~\bibnamefont {Sikivie}},
  \bibinfo {author} {\bibfnamefont {N.~S.}\ \bibnamefont {Sullivan}}, \ and\
  \bibinfo {author} {\bibfnamefont {D.~B.}\ \bibnamefont {Tanner}},\ }\href
  {\doibase 10.1103/PhysRevD.42.1297} {\bibfield  {journal} {\bibinfo
  {journal} {Phys. Rev.}\ }\textbf {\bibinfo {volume} {D42}},\ \bibinfo {pages}
  {1297} (\bibinfo {year} {1990})}\BibitemShut {NoStop}%
%%CITATION = PHRVA,D42,1297;%%
\bibitem [{\citenamefont {Hagmann}\ \emph {et~al.}(1998)\citenamefont {Hagmann}
  \emph {et~al.}}]{Hagmann:1998cb}%
  \BibitemOpen
  \bibfield  {author} {\bibinfo {author} {\bibfnamefont {C.}~\bibnamefont
  {Hagmann}} \emph {et~al.} (\bibinfo {collaboration} {ADMX}),\ }\href
  {\doibase 10.1103/PhysRevLett.80.2043} {\bibfield  {journal} {\bibinfo
  {journal} {Phys. Rev. Lett.}\ }\textbf {\bibinfo {volume} {80}},\ \bibinfo
  {pages} {2043} (\bibinfo {year} {1998})},\ \Eprint
  {http://arxiv.org/abs/astro-ph/9801286} {arXiv:astro-ph/9801286 [astro-ph]}
  \BibitemShut {NoStop}%
%%CITATION = ASTRO-PH/9801286;%%
\bibitem [{\citenamefont {Asztalos}\ \emph {et~al.}(2001)\citenamefont
  {Asztalos} \emph {et~al.}}]{Asztalos:2001tf}%
  \BibitemOpen
  \bibfield  {author} {\bibinfo {author} {\bibfnamefont {S.~J.}\ \bibnamefont
  {Asztalos}} \emph {et~al.} (\bibinfo {collaboration} {ADMX}),\ }\href
  {\doibase 10.1103/PhysRevD.64.092003} {\bibfield  {journal} {\bibinfo
  {journal} {Phys. Rev.}\ }\textbf {\bibinfo {volume} {D64}},\ \bibinfo {pages}
  {092003} (\bibinfo {year} {2001})}\BibitemShut {NoStop}%
%%CITATION = PHRVA,D64,092003;%%
\bibitem [{\citenamefont {Du}\ \emph {et~al.}(2018)\citenamefont {Du} \emph
  {et~al.}}]{Du:2018uak}%
  \BibitemOpen
  \bibfield  {author} {\bibinfo {author} {\bibfnamefont {N.}~\bibnamefont {Du}}
  \emph {et~al.} (\bibinfo {collaboration} {ADMX}),\ }\href {\doibase
  10.1103/PhysRevLett.120.151301} {\bibfield  {journal} {\bibinfo  {journal}
  {Phys. Rev. Lett.}\ }\textbf {\bibinfo {volume} {120}},\ \bibinfo {pages}
  {151301} (\bibinfo {year} {2018})},\ \Eprint
  {http://arxiv.org/abs/1804.05750} {arXiv:1804.05750 [hep-ex]} \BibitemShut
  {NoStop}%
%%CITATION = ARXIV:1804.05750;%%
\bibitem [{\citenamefont {Zhong}\ \emph {et~al.}(2018)\citenamefont {Zhong}
  \emph {et~al.}}]{Zhong:2018rsr}%
  \BibitemOpen
  \bibfield  {author} {\bibinfo {author} {\bibfnamefont {L.}~\bibnamefont
  {Zhong}} \emph {et~al.} (\bibinfo {collaboration} {HAYSTAC}),\ }\href
  {\doibase 10.1103/PhysRevD.97.092001} {\bibfield  {journal} {\bibinfo
  {journal} {Phys. Rev.}\ }\textbf {\bibinfo {volume} {D97}},\ \bibinfo {pages}
  {092001} (\bibinfo {year} {2018})},\ \Eprint
  {http://arxiv.org/abs/1803.03690} {arXiv:1803.03690 [hep-ex]} \BibitemShut
  {NoStop}%
%%CITATION = ARXIV:1803.03690;%%
\bibitem [{\citenamefont {Irastorza}\ \emph {et~al.}(2013)\citenamefont
  {Irastorza} \emph {et~al.}}]{Irastorza:2013dav}%
  \BibitemOpen
  \bibfield  {author} {\bibinfo {author} {\bibfnamefont {I.}~\bibnamefont
  {Irastorza}} \emph {et~al.} (\bibinfo {collaboration} {IAXO}),\ }\href@noop
  {} {\  (\bibinfo {year} {2013})}\BibitemShut {NoStop}%
%%CITATION = CERN-SPSC-2013-022;%%
\bibitem [{\citenamefont {Bähre}\ \emph {et~al.}(2013)\citenamefont {Bähre}
  \emph {et~al.}}]{Bahre:2013ywa}%
  \BibitemOpen
  \bibfield  {author} {\bibinfo {author} {\bibfnamefont {R.}~\bibnamefont
  {Bähre}} \emph {et~al.},\ }\href {\doibase 10.1088/1748-0221/8/09/T09001}
  {\bibfield  {journal} {\bibinfo  {journal} {JINST}\ }\textbf {\bibinfo
  {volume} {8}},\ \bibinfo {pages} {T09001} (\bibinfo {year} {2013})},\ \Eprint
  {http://arxiv.org/abs/1302.5647} {arXiv:1302.5647 [physics.ins-det]}
  \BibitemShut {NoStop}%
%%CITATION = ARXIV:1302.5647;%%
\bibitem [{\citenamefont {Melcón}\ \emph {et~al.}(2018)\citenamefont {Melcón}
  \emph {et~al.}}]{Melcon:2018dba}%
  \BibitemOpen
  \bibfield  {author} {\bibinfo {author} {\bibfnamefont {A.~Ã.}\ \bibnamefont
  {Melcón}} \emph {et~al.},\ }\href@noop {} {\  (\bibinfo {year} {2018})},\
  \Eprint {http://arxiv.org/abs/1803.01243} {arXiv:1803.01243 [hep-ex]}
  \BibitemShut {NoStop}%
%%CITATION = ARXIV:1803.01243;%%
\bibitem [{\citenamefont {Klaer}\ and\ \citenamefont
  {Moore}(2017)}]{Klaer:2017ond}%
  \BibitemOpen
  \bibfield  {author} {\bibinfo {author} {\bibfnamefont {V.~B.}\ \bibnamefont
  {Klaer}}\ and\ \bibinfo {author} {\bibfnamefont {G.~D.}\ \bibnamefont
  {Moore}},\ }\href {\doibase 10.1088/1475-7516/2017/11/049} {\bibfield
  {journal} {\bibinfo  {journal} {JCAP}\ }\textbf {\bibinfo {volume} {1711}},\
  \bibinfo {pages} {049} (\bibinfo {year} {2017})},\ \Eprint
  {http://arxiv.org/abs/1708.07521} {arXiv:1708.07521 [hep-ph]} \BibitemShut
  {NoStop}%
%%CITATION = ARXIV:1708.07521;%%
\bibitem [{\citenamefont {Graham}\ \emph {et~al.}(2015)\citenamefont {Graham},
  \citenamefont {Irastorza}, \citenamefont {Lamoreaux}, \citenamefont
  {Lindner},\ and\ \citenamefont {van Bibber}}]{Graham:2015ouw}%
  \BibitemOpen
  \bibfield  {author} {\bibinfo {author} {\bibfnamefont {P.~W.}\ \bibnamefont
  {Graham}}, \bibinfo {author} {\bibfnamefont {I.~G.}\ \bibnamefont
  {Irastorza}}, \bibinfo {author} {\bibfnamefont {S.~K.}\ \bibnamefont
  {Lamoreaux}}, \bibinfo {author} {\bibfnamefont {A.}~\bibnamefont {Lindner}},
  \ and\ \bibinfo {author} {\bibfnamefont {K.~A.}\ \bibnamefont {van Bibber}},\
  }\href {\doibase 10.1146/annurev-nucl-102014-022120} {\bibfield  {journal}
  {\bibinfo  {journal} {Ann. Rev. Nucl. Part. Sci.}\ }\textbf {\bibinfo
  {volume} {65}},\ \bibinfo {pages} {485} (\bibinfo {year} {2015})},\ \Eprint
  {http://arxiv.org/abs/1602.00039} {arXiv:1602.00039 [hep-ex]} \BibitemShut
  {NoStop}%
%%CITATION = ARXIV:1602.00039;%%
\bibitem [{\citenamefont {Koposov}\ \emph {et~al.}(2015)\citenamefont
  {Koposov}, \citenamefont {Belokurov}, \citenamefont {Torrealba},\ and\
  \citenamefont {Evans}}]{Koposov:2015cua}%
  \BibitemOpen
  \bibfield  {author} {\bibinfo {author} {\bibfnamefont {S.~E.}\ \bibnamefont
  {Koposov}}, \bibinfo {author} {\bibfnamefont {V.}~\bibnamefont {Belokurov}},
  \bibinfo {author} {\bibfnamefont {G.}~\bibnamefont {Torrealba}}, \ and\
  \bibinfo {author} {\bibfnamefont {N.~W.}\ \bibnamefont {Evans}},\ }\href
  {\doibase 10.1088/0004-637X/805/2/130} {\bibfield  {journal} {\bibinfo
  {journal} {Astrophys. J.}\ }\textbf {\bibinfo {volume} {805}},\ \bibinfo
  {pages} {130} (\bibinfo {year} {2015})},\ \Eprint
  {http://arxiv.org/abs/1503.02079} {arXiv:1503.02079 [astro-ph.GA]}
  \BibitemShut {NoStop}%
%%CITATION = ARXIV:1503.02079;%%
\bibitem [{\citenamefont {Bechtol}\ \emph {et~al.}(2015)\citenamefont {Bechtol}
  \emph {et~al.}}]{Bechtol:2015cbp}%
  \BibitemOpen
  \bibfield  {author} {\bibinfo {author} {\bibfnamefont {K.}~\bibnamefont
  {Bechtol}} \emph {et~al.} (\bibinfo {collaboration} {DES}),\ }\href {\doibase
  10.1088/0004-637X/807/1/50} {\bibfield  {journal} {\bibinfo  {journal}
  {Astrophys. J.}\ }\textbf {\bibinfo {volume} {807}},\ \bibinfo {pages} {50}
  (\bibinfo {year} {2015})},\ \Eprint {http://arxiv.org/abs/1503.02584}
  {arXiv:1503.02584 [astro-ph.GA]} \BibitemShut {NoStop}%
%%CITATION = ARXIV:1503.02584;%%
\bibitem [{\citenamefont {Drlica-Wagner}\ \emph {et~al.}(2015)\citenamefont
  {Drlica-Wagner} \emph {et~al.}}]{Drlica-Wagner:2015ufc}%
  \BibitemOpen
  \bibfield  {author} {\bibinfo {author} {\bibfnamefont {A.}~\bibnamefont
  {Drlica-Wagner}} \emph {et~al.} (\bibinfo {collaboration} {DES}),\ }\href
  {\doibase 10.1088/0004-637X/813/2/109} {\bibfield  {journal} {\bibinfo
  {journal} {Astrophys. J.}\ }\textbf {\bibinfo {volume} {813}},\ \bibinfo
  {pages} {109} (\bibinfo {year} {2015})},\ \Eprint
  {http://arxiv.org/abs/1508.03622} {arXiv:1508.03622 [astro-ph.GA]}
  \BibitemShut {NoStop}%
%%CITATION = ARXIV:1508.03622;%%
\bibitem [{\citenamefont {{Massari}}\ and\ \citenamefont
  {{Helmi}}(2018)}]{2018arXiv180501839M}%
  \BibitemOpen
  \bibfield  {author} {\bibinfo {author} {\bibfnamefont {D.}~\bibnamefont
  {{Massari}}}\ and\ \bibinfo {author} {\bibfnamefont {A.}~\bibnamefont
  {{Helmi}}},\ }\href@noop {} {\bibfield  {journal} {\bibinfo  {journal} {ArXiv
  e-prints}\ } (\bibinfo {year} {2018})},\ \Eprint
  {http://arxiv.org/abs/1805.01839} {arXiv:1805.01839} \BibitemShut {NoStop}%
\bibitem [{\citenamefont {Ivezic}\ \emph {et~al.}(2008)\citenamefont {Ivezic},
  \citenamefont {Tyson}, \citenamefont {Allsman}, \citenamefont {Andrew},\ and\
  \citenamefont {Angel}}]{Ivezic:2008fe}%
  \BibitemOpen
  \bibfield  {author} {\bibinfo {author} {\bibfnamefont {Z.}~\bibnamefont
  {Ivezic}}, \bibinfo {author} {\bibfnamefont {J.~A.}\ \bibnamefont {Tyson}},
  \bibinfo {author} {\bibfnamefont {R.}~\bibnamefont {Allsman}}, \bibinfo
  {author} {\bibfnamefont {J.}~\bibnamefont {Andrew}}, \ and\ \bibinfo {author}
  {\bibfnamefont {R.}~\bibnamefont {Angel}} (\bibinfo {collaboration} {LSST}),\
  }\href@noop {} {\  (\bibinfo {year} {2008})},\ \Eprint
  {http://arxiv.org/abs/0805.2366} {arXiv:0805.2366 [astro-ph]} \BibitemShut
  {NoStop}%
%%CITATION = ARXIV:0805.2366;%%
\bibitem [{\citenamefont {Consiglio}\ \emph {et~al.}(2017)\citenamefont
  {Consiglio}, \citenamefont {Turner}, \citenamefont {Beck}, \citenamefont
  {Meier}, \citenamefont {Silich},\ and\ \citenamefont
  {Zhao}}]{0004-637X-850-1-54}%
  \BibitemOpen
  \bibfield  {author} {\bibinfo {author} {\bibfnamefont {S.~M.}\ \bibnamefont
  {Consiglio}}, \bibinfo {author} {\bibfnamefont {J.~L.}\ \bibnamefont
  {Turner}}, \bibinfo {author} {\bibfnamefont {S.}~\bibnamefont {Beck}},
  \bibinfo {author} {\bibfnamefont {D.~S.}\ \bibnamefont {Meier}}, \bibinfo
  {author} {\bibfnamefont {S.}~\bibnamefont {Silich}}, \ and\ \bibinfo {author}
  {\bibfnamefont {J.-H.}\ \bibnamefont {Zhao}},\ }\href
  {http://stacks.iop.org/0004-637X/850/i=1/a=54} {\bibfield  {journal}
  {\bibinfo  {journal} {The Astrophysical Journal}\ }\textbf {\bibinfo {volume}
  {850}},\ \bibinfo {pages} {54} (\bibinfo {year} {2017})}\BibitemShut
  {NoStop}%
\end{thebibliography}%
%%%%%%%%%%%%%%%%%%%%%%%%%%%%%%%%%%%%%%%%%%%%%%%%%%%%%

\end{document}